\documentclass[bjps, noinfoline, natbib]{imsart}
\pdfoutput=1

\makeatletter
\setattribute{copyright}{text}{\url@fmt{}{\itshape}{\journal@name}{\journal@url}\break%
                               \@ifnonempty{2017, Vol. 31, No. 4, \@pagerange\break}%
                               \@ifnonempty{\journal@issn\break}%
                               \doi@text}
\makeatother

\usepackage{hyperref}
\let\inlinecite\citet
\let\cite\relax
\def\cite*#1{\citealp{#1}}

\usepackage{array}
\newcolumntype{C}[1]{>{\centering\let\newline\\\arraybackslash\hspace{0pt}}p{#1}}
\let\tablewidth\textwidth
\newcounter{tabnotecounter}
\newcommand{\tabnoteref}[1]{\stepcounter{tabnotecounter}\textsuperscript{\scriptsize(\thetabnotecounter)}}
\newcommand{\tabnotetext}[3][]{\textsuperscript{\scriptsize#1}#3\\}

\usepackage{ragged2e}

\usepackage{microtype}

\volume{31}
\issue{4}
\pubyear{2017}
\firstpage{686}
\lastpage{691}
\doi{10.1214/17-BJPS365B}

\makeatletter
\def\journal@name{To appear in the Brazilian Journal of Probability and Statistics}
\makeatother

\setattribute{bauthor}{style}{}
\ifx \binits  \undefined \def \binits#1{#1}\fi
\ifx \bauthor  \undefined \def \bauthor#1{\textrm{#1}}\fi
\ifx \batitle  \undefined \def \batitle#1{#1}\fi
\ifx \bjtitle  \undefined \def \bjtitle#1{\textit{#1}}\fi
\ifx \byear  \undefined \def \byear#1{#1}\fi
\ifx \bfpage  \undefined \def \bfpage#1{#1}\fi
\ifx \blpage  \undefined \def \blpage #1{#1}\fi
\ifx \bctitle  \undefined \def \bctitle#1{#1}\fi
\ifx \bbtitle  \undefined \def \bbtitle#1{\textit{#1}}\fi
\ifx \bsnm \undefined \def \bsnm#1{#1}\fi
\ifx \botherref \undefined \def \botherref #1{#1}\fi
\ifx \bchapter \undefined \def \bchapter#1{#1}\fi
\ifx \oauthor \undefined \def \oauthor#1{#1}\fi
\ifx \arxivurl  \undefined \def \arxivurl#1{#1}\fi
\renewcommand{\url}[1]{\href{#1}{#1}}
\setattribute{thebibliography}{size}{\footnotesize}

\usepackage{microtype}
\usepackage{lmodern}

\begin{document}

\begin{frontmatter}

\title{\vspace*{3ex}Comment: A brief survey of the current state of play for
Bayesian computation in data science at big-data scale}
\runtitle{Comment}

\begin{aug}

\author{\fnms{David}~\snm{Draper}\thanksref{A}}
\and
\author{\fnms{Alexander}~\snm
{Terenin}\thanksref{B}}

\runauthor{D. Draper and A. Terenin}
\affiliation[A]{University of California, Santa Cruz}
\affiliation[B]{Imperial College London}

\address{Baskin School of Engineering\\University of
California\\1156 High Street\\
Santa Cruz, California 95064\\ USA\\
E-mail: \href{mailto:draper@ucsc.edu}{draper@ucsc.edu}
}
\end{aug}

\received{\smonth{6} \syear{2017}}
\accepted{\smonth{6} \syear{2017}}

\end{frontmatter}

We wish to contribute to the discussion of this interesting paper by
offering our views on the current best methods for Bayesian
computation, both at Big-Data scale and with smaller data sets, as
summarized in Table~\ref{t:methods}. This table is certainly an
over-simplification of a highly complicated area of research in
constant (present and likely future) flux, but we believe that
constructing summaries of this type is worthwhile despite their
drawbacks, if only to facilitate further discussion.

The entries in the table are tailored to problems---such as
classification and regression---in which the data set is organized
into $n$ rows (representing subjects of study, which may or may not be
organized hierarchically, for example, patients nested in hospitals)
and $p$ columns of variables measured on the subjects, but the table
may be relevant for other data structures as well (for example, data
sets in \textit{topic modeling} (e.g., \cite*{blei12}) may be
visualized as consisting of $n$ words (nested inside documents) and $p$
topics). The rows of Table~\ref{t:methods} play out the four
combinations of big and small $n$ and $p$, and the columns identify the
four major hardware configurations currently available: single CPU,
single GPU, cluster of CPUs, and cluster of GPUs. A more detailed table
would have broken both $n$ and $p$ out into (small, medium, large), and
further detail could have been supplied in a more model-specific
fashion, but in our view Table~\ref{t:methods} already makes some
interesting comparisons.

For us, ``Big Data'' refers to settings in which the available data set
is too big to store and process in memory on a single machine, or
\textit{worker} (consisting of one single-threaded CPU, some RAM, and
some disk storage). In our table this corresponds to big $n$, which in
today's environment typically equates to a data set that takes tens of
gigabytes (or more) to store. For our purposes, big $p$ is in the
hundreds of thousands, tiny $p$ is a few hundred, and small $p$ is in between.

%
\begin{table}[t]
\justify
\caption{Our view of the current best methods for simulation-based
Bayesian computation, as a function of number of observations ($n$),
number of predictors ($p$), and available hardware. Big $n$ specifies
data sets that take tens of gigabytes (or more) to store; tiny $p$ is a
few hundred; big $p$ is hundreds of thousands\hspace*{0.76\textwidth}\vspace*{1ex}}
\label{t:methods}
\def\arraystretch{1.5}
\begin{tabular*}{\tablewidth}{@{\extracolsep{\fill
}}lcC{75pt}C{75pt}C{75pt}C{60pt}@{}}
\hline
$n$ & $p$ & Single CPU & Single GPU & CPU Cluster & GPU Cluster \\
\hline
Small & Small & Standard MCMC & GPU-accelerated particle filters &
Independent parallel chains & not needed \\[16pt]
Small & Big & Hamiltonian Monte Carlo (HMC) or Gibbs sampling\tabnoteref
{tab1a} &
HMC or GPU-accelerated Gibbs sampling & Asynchronous Gibbs
sampling\tabnoteref{tab1b} &
No current Bayesian method \\[16pt]
Big & Small & Continuous-time MCMC with subsampling and control
variates\tabnoteref{tab1c} &
Continuous-time MCMC or GPU-accelerated Gibbs sampling\tabnoteref
{tab1d} &
Asynchronous Gibbs sampling\tabnoteref{tab1e} or methods based on
sharding &
No current Bayesian method\\[16pt]
Big & Big & Hopeless: point estimates only & Model-specific\tabnoteref
{tab1f} &
Model-specific\tabnoteref{tab1g} & No current Bayesian method \\
\hline
\strut
\end{tabular*}
\tabnotetext[(1)]{tab1a}{HMC if gradient is available, else Gibbs.}
\tabnotetext[(2)]{tab1b}{The exact algorithm if Metropolis--Hastings
correction is cheap enough, otherwise the approximate algorithm if
diagnostic indicates appropriate.}
\tabnotetext[(3)]{tab1c}{If gradient is available and $p$ is tiny.}
\tabnotetext[(4)]{tab1d}{ZigZag (see text) or BPS with parallel
gradient evaluations if $p$ is tiny;
GPU-accelerated Gibbs if gradient unavailable and/or $p$ is medium-size.}
\tabnotetext[(5)]{tab1e}{If data augmentation is possible with the
model under study.}
\tabnotetext[(6)]{tab1f}{May be possible for gradient-based methods in
some models (current research).}
\tabnotetext[(7)]{tab1g}{An example is topic modeling, using AD--LDA
(approximate) or P\'olya Urn LDA (essentially exact with
big $n$).}
\end{table}

Our comments on the individual cells in the table are as follows.
\begin{itemize}

\item

$( n, p ) =$ (Small, Small) is the current Bayesian computational
comfort zone: an example would be logistic regression using a
regularization prior (e.g., the horseshoe (\cite*{carvalho10})) with
an $n$ of a few thousand and $p$ on the order of 100.
\begin{itemize}

\item

Standard MCMC on a single CPU is fine here, with home-grown code or
using an environment such as \texttt{WinBUGS} (\cite*{lunn00}) or
\texttt{RJAGS} (\cite*{plummer03}).

\item

If you have a single GPU and need answers extremely quickly,
GPU-accelerated particle filters (\cite*{lee10}) are a good option
with some models.

\item

If instead you have access to a cluster of workers, you can of course
readily run independent parallel chains on each worker and merge the
results. This includes computing on a single machine with multiple
cores/threads, for example via \texttt{doParallel} or \texttt{snow} in
\texttt{R}.

\item

A cluster of GPUs would be overkill in this situation.
\vspace*{3ex}
\end{itemize}

\item

$( n, p ) =$ (Small, Big) is already at the research frontier with some
hardware configurations.
\begin{itemize}

\item

Using a single worker, you can do Hamiltonian Monte Carlo (HMC,
\cite*{betancourt17}) if the gradient of the log posterior is
available (e.g., with non-discrete likelihoods), writing the code
yourself or fitting your model in \texttt{Stan} (\cite*{stan17}), or
you can use Gibbs sampling in settings with or without a gradient.

\item

With a single GPU, HMC is again a good option with a gradient (\cite*
{beam15}), or you can use GPU-accelerated Gibbs sampling (\cite*
{terenin16}). The latter method can be better in models with
heavy-tailed error distributions, where HMC sometimes performs poorly,
but HMC within Gibbs would be even better than pure Gibbs in such problems.

\item

If you have access to a cluster of CPUs, one good method is
asynchronous Gibbs sampling (\cite*{terenin17a}), using either (a)
the exact algorithm, if the Metropolis-Hastings correction built into
the method is cheap enough or (b) the (much faster) approximate
algorithm, if the diagnostic described in the paper indicates that this
is safe.

\item

It would appear that no one knows how to make efficient fully Bayesian
use of a GPU cluster at present: the problem is that GPU computation is
massively faster than available network speed, so it's typically not at
all clear how to keep all of the GPUs busy at once.
\end{itemize}

\item

$( n, p ) =$ (Big, Small) is an active area of recent and current research.
\begin{itemize}

\item

On a single CPU, in our view the most promising approach with big $n$
and tiny $p$, when a gradient is available, involves methods based on
creating continuous-time stochastic processes without discretization
that sample correctly from the posterior of interest---these include
ScaLE (\cite*{pollock16}), ZigZag sampling (\cite*{bierkens16}),
and the Bouncy Particle Sampler (BPS, \cite*{bouchardcote15}). The
key advantage of these methods is the ability to run while evaluating
one data point at a time.

\item

With a single GPU, two good options are (a) BPS (with parallel gradient
evaluations) if $p$ is tiny and (b) GPU-accelerated Gibbs sampling if
the gradient is unavailable and/or if $p$ is medium in size (up to
10,000, say).

\item

A cluster of workers at your disposal lands you in the cell into which
Steve Scott's present paper fits, under \textit{divide-and-conquer}
methods based on sharding---other methods that partition the data set
into small subsets of subjects, make standard calculations on each
small data set, and then combine at the end include a variety of
techniques developed by David Dunson and his collaborators (e.g.,
\cite*{srivastiva15}). Many of the sharding papers concentrate on
examples in which $p$ is tiny; Asynchronous Gibbs sampling is a good
alternative with larger $p$ when data augmentation is possible in the
model at hand. One of the issues that sharding-based methods face with
larger $p$ is ensuring convergence on each of the individual shards.

\item

With a cluster of GPUs, as noted above, the profession is currently at
a loss to make fully effective use of the hardware for posterior
exploration (as opposed to optimization: see below).
\end{itemize}

\item

$( n, p ) =$ (Big, Big) is where current methods start to reveal their
(serious) limitations.
\begin{itemize}

\item

Using only a single CPU, fully Bayesian posterior exploration in this
case is hopeless. The best you can hope for is employing
optimization-based methods
to obtain the maximum a posteriori (MAP) estimate. As a large body of
work in machine learning shows, this is sometimes adequate to solve the
problem at hand.

\item

With a single GPU, success becomes model-specific. It may be possible
to adapt gradient-based methods to drive a GPU efficiently in some
models---this is a topic of current research.

\item

Employing a cluster of workers, with $n$ and $p$ both big, good results
are currently also available only on a model-specific basis---topic
models provide an example, with Approximate Distributed LDA (AD--LDA,
\cite*{newman09}) offering approximate results and P\'olya Urn LDA
(\cite*{terenin17b}) producing answers that are exact up to Monte
Carlo noise with big $n$. If it's sufficient to obtain MAP estimates in
the problem you're working on, by far the best Big-Data optimization
algorithm at present is \textit{stochastic gradient descent} (SGD; see
\cite*{recht11} for an asynchronous parallellized version of the
method). One of the great advantages of SGD is that it can be coded in
such a way that the data set can be \textit{streamed} through a
processor, without any need to store it---see \inlinecite{tran16} for
an \texttt{R} implementation for some models in the \texttt{CRAN}
package \texttt{sgd}.

It's worth noting that ScaLE, ZigZag, and BPS can all handle streaming
data, producing accurate full-Bayes posterior exploration (i.e., not
just MAP estimates) in the ($n$ big, $p$ small) region of problem
space. This is in contrast to other methods such as \textit{stochastic
gradient Langevin dynamics} (SGLD, \cite*{welling11}), which
explores the posterior so slowly that it might as well just be an
optimization method. SGLD provides a vivid example of the meta-theorem
that if you have a simulation-based method that has an asymptotic
convergence proof but that mixes extremely poorly, the convergence
proof may be useless as a guide to the practical effectiveness of the method.

\item

And finally, once again, nobody knows how to make efficient fully
Bayesian use of a cluster of GPUs in the $( n, p ) =$ (Big, Big)
setting, where such a cluster would be most needed if it could be
effectively utilized. If it's sufficient in your problem to settle for
MAP estimates, see \inlinecite{barkalov16} for one approach to computing
them on a GPU cluster.
\end{itemize}
\end{itemize}

Three more concluding comments:
\begin{itemize}

\item

At several points we said things like ``if you have access to [hardware
configuration $X$],'' but we wish to emphasize that \textit{everybody}
currently has access to the hardware in all four columns of Table~\ref{t:methods}---including people in academia, not just in industry---by
renting this hardware from any of a number of cloud-computing
companies, typically for less than US\$1 per hour (and universities can
often persuade such companies to donate many hours in the cloud to them
for free). If you're currently using only the first column of Table~\ref{t:methods} as your hardware resource and you want to begin exploring
the Big-Data world, we would encourage you to venture into columns 2
and 3. If you do so, and you can handle the programming (see below),
you'll find, as we have, that for many problems a single GPU is better
than 100 CPUs.

\item

Having just issued an invitation to explore new hardware, it must be
admitted that the availability of user-friendly software to make best
use of GPUs and CPU clusters is currently nowhere near the level of
ease of use of, for example, \texttt{CRAN} packages in \texttt{R}. One
promising recent development to note, however, is that at least one
hardware manufacturer has recently introduced an external GPU node that
makes it possible to drive a GPU with \texttt{CUDA} programs residing
on your laptop or desktop---this brings Bayes on GPUs one step
further away from the research frontier and closer to day-to-day applications.

\item

One further note on streaming, GPUs, and the value of contemporary
hardware: to work efficiently on a GPU, you either have to (1) hold
your data set in the GPU's memory or (2) stream your data (if it's too
big to fit in memory). Option (2) is how the machine-learning community
has obtained its remarkable recent successes with the use of \textit
{deep learning} (e.g., \cite*{schmidhuber15}) via SGD on a GPU to
solve the image classification problem (\cite*{krizhevsky12}).

This brings us to our final meta-theorem: statisticians run the serious
risk of being marginalized in the field of data science, ceding the
high ground to machine learning when, in our view, this field should
involve an equal partnership between mathematical sciences (applied
mathematics, statistics), computing sciences (machine learning,
database organization and management, the hardware-software interface),
and subject-matter expertise in the problem at hand.
\end{itemize}

We thank Steve for a stimulating paper, and we would be interested in
his comments on how sharding methods in general, and his approach in
particular, fit into the framework of our table, both currently and in
the future.


%


\end{document}